 \documentclass[showpacs,aps,floatfix,prd,11pt,superscriptaddress]{revtex4}

\usepackage{graphicx}
\usepackage{dcolumn}
\usepackage{amsmath}
\usepackage{epsfig}
\usepackage{hhline}
\usepackage{ifthen}

\newboolean{istwocolumn}
  \setboolean{istwocolumn}{false}

\RequirePackage{xspace}

\usepackage{relsize}
\def\babar{\mbox{\slshape B\kern-0.1em{\smaller A}\kern-0.1em
    B\kern-0.1em{\smaller A\kern-0.2em R}}}

\def\Kbar  {\kern 0.2em\overline{\kern -0.2em K}{}\xspace}

\def\Kz    {\ensuremath{K^0}\xspace}
\def\Kzb   {\ensuremath{\Kbar^0}\xspace}
\def\KzKzb {\ensuremath{\Kz \kern -0.16em \Kzb}\xspace}
\def\Kp    {\ensuremath{K^+}\xspace}
\def\Km    {\ensuremath{K^-}\xspace}

\def\KpKm  {\ensuremath{\Kp \kern -0.16em \Km}\xspace}

\def\Dbar    {\kern 0.2em\overline{\kern -0.2em D}{}\xspace}

\def\Dz      {\ensuremath{D^0}\xspace}
\def\Dzb     {\ensuremath{\Dbar^0}\xspace}
\def\DzDzb   {\ensuremath{\Dz {\kern -0.16em \Dzb}}\xspace}
\def\Dp      {\ensuremath{D^+}\xspace}
\def\Dm      {\ensuremath{D^-}\xspace}

\def\DpDm    {\ensuremath{\Dp {\kern -0.16em \Dm}}\xspace}

\def\B       {\ensuremath{B}\xspace}
\def\Bbar    {\kern 0.18em\overline{\kern -0.18em B}{}\xspace}

\def\Bz      {\ensuremath{B^0}\xspace}
\def\Bzb     {\ensuremath{\Bbar^0}\xspace}
\def\BzBzb   {\ensuremath{\Bz {\kern -0.16em \Bzb}}\xspace}
\def\Bu      {\ensuremath{B^+}\xspace}
\def\Bub     {\ensuremath{B^-}\xspace}

\def\BpBm    {\ensuremath{\Bu {\kern -0.16em \Bub}}\xspace}

\mathchardef\Upsilon="7107
\def\Y#1S{\ensuremath{\Upsilon{(#1S)}}\xspace}

\mathchardef\Deltares="7101
\mathchardef\Xi="7104
\mathchardef\Lambda="7103
\mathchardef\Sigma="7106
\mathchardef\Omega="710A

\def\Deltabar{\kern 0.25em\overline{\kern -0.25em \Deltares}{}\xspace}
\def\Lbar{\kern 0.2em\overline{\kern -0.2em\Lambda\kern 0.05em}\kern-0.05em{}\xspace}
\def\Sigbar{\kern 0.2em\overline{\kern -0.2em \Sigma}{}\xspace}
\def\Xibar{\kern 0.2em\overline{\kern -0.2em \Xi}{}\xspace}
\def\Obar{\kern 0.2em\overline{\kern -0.2em \Omega}{}\xspace}
\def\Nbar{\kern 0.2em\overline{\kern -0.2em N}{}\xspace}
\def\Xb{\kern 0.2em\overline{\kern -0.2em X}{}\xspace}

\newcommand{\tev}{\ensuremath{\mathrm{\,Te\kern -0.1em V}}\xspace}
\newcommand{\gev}{\ensuremath{\mathrm{\,Ge\kern -0.1em V}}\xspace}
\newcommand{\mev}{\ensuremath{\mathrm{\,Me\kern -0.1em V}}\xspace}
\newcommand{\kev}{\ensuremath{\mathrm{\,ke\kern -0.1em V}}\xspace}
\newcommand{\ev}{\ensuremath{\mathrm{\,e\kern -0.1em V}}\xspace}
\newcommand{\gevc}{\ensuremath{{\mathrm{\,Ge\kern -0.1em V\!/}c}}\xspace}
\newcommand{\mevc}{\ensuremath{{\mathrm{\,Me\kern -0.1em V\!/}c}}\xspace}
\newcommand{\gevcc}{\ensuremath{{\mathrm{\,Ge\kern -0.1em V\!/}c^2}}\xspace}
\newcommand{\mevcc}{\ensuremath{{\mathrm{\,Me\kern -0.1em V\!/}c^2}}\xspace}

\def\mus  {\ensuremath{\rm \,\mus}\xspace}

\def\mus        {\ensuremath{\,\mu{\rm s}}\xspace}    

\def\calA{{\ensuremath{\cal A}}\xspace}

\def\pep2{PEP-II}

\def\gsim{{~\raise.15em\hbox{$>$}\kern-.85em
          \lower.35em\hbox{$\sim$}~}\xspace}
\def\lsim{{~\raise.15em\hbox{$<$}\kern-.85em
          \lower.35em\hbox{$\sim$}~}\xspace}

\def\CP                {\ensuremath{C\!P}\xspace}

\xspace

\newcommand{\progtp}    [1]  {{Prog.\ Th.\ Phys.\ {\bf #1}}}

\def\jetset74   {\mbox{\tt Jetset \hspace{-0.5em}7.\hspace{-0.2em}4}\xspace}

\reversemarginpar
\begin{document}


\title{ \Large Impact of tag-side interference on time-dependent $CP$ asymmetry
               measurements using coherent \BzBzb pairs}

\author{Owen Long}
\affiliation{University of California at Santa Barbara, Santa Barbara, CA 93106, USA }
\author{Max Baak}
\affiliation{NIKHEF, National Institute for Nuclear and High Energy Physics, 1009 DB Amsterdam, The Netherlands}
\author{Robert N. Cahn}
\affiliation{Lawrence Berkeley National Laboratory and University of California, Berkeley, CA 94720, USA }
\author{David Kirkby}
\affiliation{University of California at Irvine, Irvine, CA 92697, USA }

\date{\today}

\begin{abstract}
  Interference between CKM-favored $b\rightarrow c\overline{u}d$ and doubly-CKM-suppressed
  $\overline{b}\rightarrow \overline{u} c \overline{d}$ amplitudes in final states used
  for \B flavor tagging gives deviations from the standard time evolution assumed in $CP$-violation
  measurements at $B$ factories producing coherent \BzBzb pairs.
  We evaluate these deviations for the standard time-dependent $CP$-violation
  measurements, the uncertainties they introduce in the measured quantities,
  and give suggestions for minimizing them.
  The uncertainty in the measured $CP$ asymmetry for $CP$ eigenstates is $\approx 2$\%
  or less.
  The time-dependent analysis of $D^*\pi$, proposed for measuring $\sin(2\beta+\gamma)$,
  must incorporate possible tag-side interference, which could produce asymmetries as large
  as the expected
  signal asymmetry.
\end{abstract}


\maketitle


\section{Introduction}

Measurements of time-dependent $CP$ asymmetries in \Bz decays provide information about the
irreducible phase contained in the Cabibbo-Kobayashi-Maskawa (CKM)
quark-mixing matrix~\cite{CKM}, which describes $CP$
violation in the Standard Model.
If a specific $B$ decay final state has contributions from
more than one amplitude
and these amplitudes have different $CP$-violating weak phases,
interference can produce a non-zero $CP$ asymmetry.
An essential ingredient in $CP$ violation measurements in \Bz decays is
flavor tagging.
In this paper, we point out a subtlety of flavor tagging that has
been overlooked or ignored in most recent $CP$ violation analyses,
describe the impact of this omission, and propose how to address it
in some future measurements.

In the current asymmetric $B$-factories~\cite{oddone}, PEP-II and KEKB,
\BzBzb meson pairs are produced in $e^{+}e^{-}$ interactions at the $\Upsilon$(4S) resonance,
where the pair evolves coherently in a $P$-wave state until one of the $B$ mesons decays.
Typically, one $B$ decay is fully reconstructed and the
flavor (whether it's a \Bz or \Bzb)
of this $B$, at the time of the other $B$'s decay,
is inferred from the decay products of the other $B$ (the tag $B$).
At the time of the tag $B$ meson decay, the $B$ mesons are known to be in opposite flavor states.
In terms of the time difference between the two $B$ decays, $\Delta t \equiv t_{\rm rec} - t_{\rm tag}$,
the time-dependent $CP$ asymmetry is defined as
\begin{equation}
   A_{CP}( \Delta t ) \equiv
      \frac{ N\left({\rm tag}\ \Bz,\Delta t\right)  -  N\left({\rm tag}\ \Bzb,\Delta t\right) }
           { N\left({\rm tag}\ \Bz,\Delta t\right)  +  N\left({\rm tag}\ \Bzb,\Delta t\right) }\,,
\end{equation}
where $N$ is the number of events at $\Delta t$ with a \Bz or \Bzb as the tag $B$.

Charged leptons and kaons are often used to infer the flavor of the tag $B$ meson.
The charge of a lepton from a semi-leptonic $B$ decay has the same sign
as the charge of the $b$ quark that produced it.
For example, a high-momentum $e^+$ ($e^-$) would indicate that the tag $B$
was a \Bz (\Bzb) at the time of its decay.
Similarly, a $K^+$($K^-$) more often than not comes from a \Bz(\Bzb).
This works because the most likely $b$ decay is $b\rightarrow c$
and the most likely $c$ decay is $c\rightarrow s$; thus the $s$
quark usually has the same charge as the $b$ quark.
The lepton or kaon charge does not always correctly indicate the tag-$B$ flavor.
Mistags can come from incorrect particle identification or
other $B$ decay chains that produce wrong-sign leptons or kaons.
The mistag fraction must be measured in order to determine the true
$CP$ asymmetry from the measured one.

It is usually assumed that the measured $CP$ asymmetry
is entirely due to the interfering amplitudes contributing to the fully
reconstructed $B$ decay mode, and that the individual tagging states,
such as $\Bzb\rightarrow D^+\pi^-$, are dominated by a single $B$ decay amplitude.
In other words, if only one $B$ decay amplitude contributes to the
tagging final state, it is safe to assume that all interference effects,
such as CP violation, are due to the evolution of the fully reconstructed $B$.
This assumption, which is valid for semi-leptonic $B$ decays,
ignores the possibility of suppressed contributions
to the tag-side final state with different weak phases,
such as happens for non-leptonic decays.

\begin{figure}[!h]
\begin{center}%
\ifthenelse{\boolean{istwocolumn}}{
  \epsfig{figure=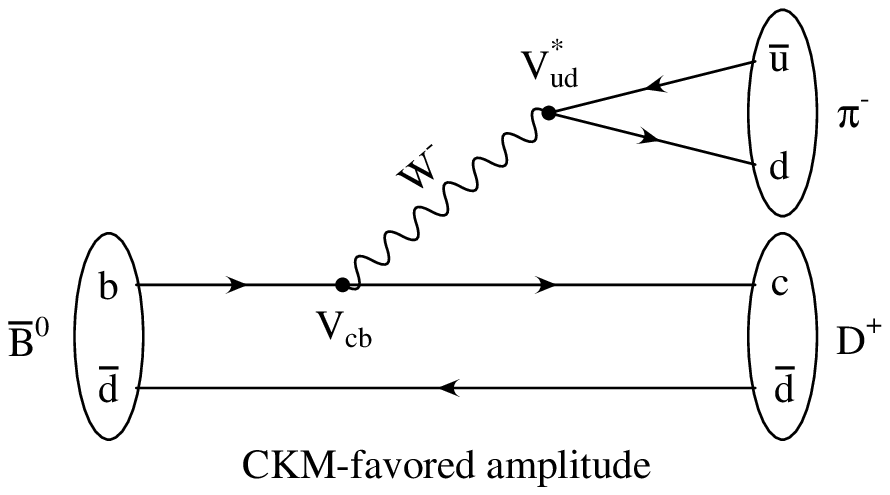,width=0.9\linewidth}
  \epsfig{figure=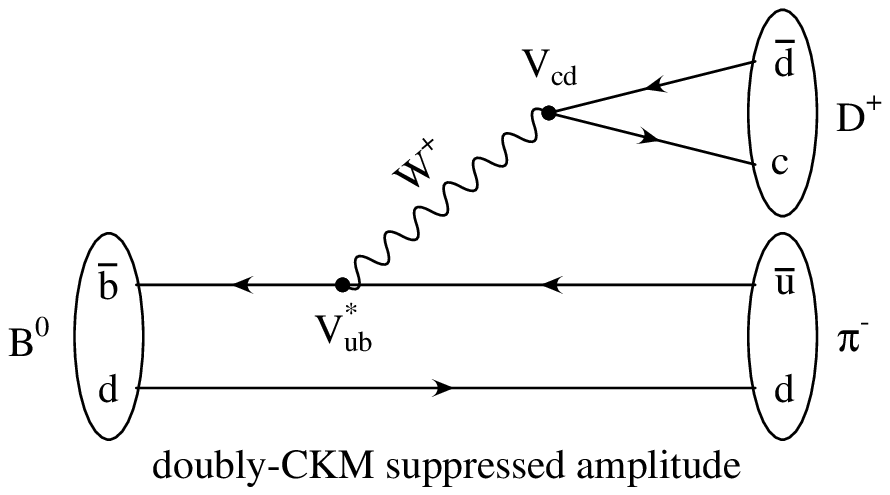,width=0.9\linewidth}
}{
  \epsfig{figure=fig1a.eps,width=0.45\linewidth}
  \epsfig{figure=fig1b.eps,width=0.45\linewidth}
}
\caption{The CKM-favored amplitude (left) and doubly-CKM-suppressed amplitude
         (right) for the final state $D^+\pi^-$. With respect to the dominant
         contribution, the latter is suppressed by the approximate
         ratio $|(V_{ub}^* V_{cd}) / (V_{cb} V_{ud}^*)| \approx 0.02$
         and has a relative weak phase difference of $\gamma$.}
\label{fig:favsub}
\end{center}
\end{figure}

These suppressed contributions may be important for kaon tags.
For example, the $D^+\pi^-$ final state with $D^+\rightarrow K^-\pi^+\pi^+$
, which is usually associated with a $\Bzb$ decay, can also be reached from a \Bz
through a $\overline{b}\rightarrow \overline{u} c \overline{d}$ decay.
Its amplitude is suppressed relative to the dominant $\Bzb$
decay amplitude ($b\rightarrow c\overline{u}d$) by a factor of roughly
$|(V_{ub}^* V_{cd}^{\vphantom{*}}) / (V_{cb}^{\vphantom{*}} V_{ud}^*)| \approx 0.02$, and has a
relative weak phase difference of $\gamma$.
Both Feynman diagrams are shown in Fig.~\ref{fig:favsub}.
The tag-side $b\rightarrow c\overline{u}d$ and $\overline{b}\rightarrow \overline{u} c \overline{d}$
amplitudes interfere, and, through the coherent evolution of the \BzBzb pair,
alter the time evolution of $A_{CP}(\Delta t)$.
The subject of this paper is to investigate the consequences of
this small tag-side interference in some of the standard time-dependent $CP$-asymmetry measurements at 
$B$ factories that use coherent $B$ decays.

In Sections~\ref{sec:formalism} -- \ref{sec:rprime-and-deltaprime},
we review the general formalism for
describing the coherent evolution of the \BzBzb system,
define our notation for describing the tag-side amplitude,
and state the assumptions we employ in our analysis.
In Section~\ref{sec:mistag}, we evaluate how tag-side interference
affects the mistag fraction measured from the amplitude
of the time-dependent mixing (not $CP$) asymmetry.
We find that the tag-side interference effects are {\em not}
simply absorbed into the mistag fractions and that, to first order,
the mistag fractions are unchanged by tag-side interference.
In Section~\ref{sec:sin2beta}, we evaluate
the uncertainty, due to tag-side interference, in the standard
mixing-induced $CP$ asymmetry measurements --
$\sin 2\beta$ from $J/\psi K_s$ and the $CP$ asymmetry in
$\pi^+\pi^-$.
%
%
We find that the uncertainties are at most 5\%, in the
most conservative estimation, and can be limited to
$<2\%$ in most cases with reasonable assumptions.
Finally, in Section~\ref{sec:sin2bpg}, we evaluate how tag-side interference 
affects some of the time-dependent techniques that have
been proposed for measuring $\gamma$ (e.g. the time-dependent
analysis of $D^{*+}\pi^-$).
Here, we find that tag-side interference effects can be as large as
the signal asymmetry.
We propose a technique for performing the analysis in a general
way, which does not require assumptions about the size of tag-side interference
effects and maximizes the statistical sensitivity to $(2\beta+\gamma)$.
We summarize our conclusions in Section~\ref{sec:conclusions}.

\section{General Coherent Formalism}
\label{sec:formalism}

In this section, we define our formalism for describing
the time evolution of a pair of neutral $B$ mesons 
that are coherently produced in an $\Upsilon(4S)$ decay and then
subsequently decay to arbitrary final states $f_{\rm t}$ and $f_{\rm r}$ at times
$t_{\rm t}$ and $t_{\rm r}$, respectively, measured in the parent B meson's rest frame.
The ``t'' (``r'') subscript refers to the tag (reconstructed) $B$ meson or its final state.
The amplitude for this process is proportional to
%
%
\ifthenelse{\boolean{istwocolumn}}{
%
   \begin{eqnarray}
   \calA & = &
   \langle f_{\rm t} | \Bz_{\text{phys}}(t_{\rm t})\rangle
   \langle f_{\rm r} | \Bzb_{\text{phys}}(t_{\rm r})\rangle \nonumber \\
 & - &
   \langle f_{\rm t} | \Bzb_{\text{phys}}(t_{\rm t})\rangle
   \langle f_{\rm r} | \Bz_{\text{phys}}(t_{\rm r})\rangle
   \,,
   \end{eqnarray}
}{
%
  \begin{equation}
  \calA =
  \langle f_{\rm t} | \Bz_{\text{phys}}(t_{\rm t})\rangle
  \langle f_{\rm r} | \Bzb_{\text{phys}}(t_{\rm r})\rangle -
  \langle f_{\rm t} | \Bzb_{\text{phys}}(t_{\rm t})\rangle
  \langle f_{\rm r} | \Bz_{\text{phys}}(t_{\rm r})\rangle
  \,,
  \end{equation}
}
where $\Bz_{\text{phys}}(t)$ ($\Bzb_{\text{phys}}(t)$) denotes an
initially-pure \Bz (\Bzb) state after a time $t$. 
The relative minus sign between the terms reflects the antisymmetry of the $P$-wave 
\BzBzb state.
Integrating over all
directions for either \B and the experimentally-unobservable average
decay time $(t_{\rm t}+t_{\rm r})/2$, we obtain a corresponding decay rate
proportional to ($\Delta t\equiv t_{\rm r}-t_{\rm t}$)
\begin{equation}
F(\Delta t) =
e^{-\Gamma |\Delta t|}\,\left|
a_{+} g_+(\Delta t) + a_{-} g_-(\Delta t)
\right|^2
\,,
\label{eq:F12-defn}
\end{equation}
where $\Gamma$ is the average neutral B eigenstate decay rate and we define
\ifthenelse{\boolean{istwocolumn}}{
  \begin{equation}
  \begin{split}
  g_{\pm}(\Delta t)\equiv \frac{1}{2}\Big( \
  &e^{-i\Delta m \Delta t/2} e^{-\Delta\Gamma\Delta t/4}  \\
 \pm \ \ &e^{+i\Delta m \Delta t/2} e^{+\Delta\Gamma\Delta t/4} \ \Big)
  \label{eq:g-defns}
  \end{split}
  \end{equation}
}{
  \begin{equation}
  g_{\pm}(\Delta t)\equiv \frac{1}{2}\left(
  e^{-i\Delta m \Delta t/2} e^{-\Delta\Gamma\Delta t/4} \pm
  e^{+i\Delta m \Delta t/2} e^{+\Delta\Gamma\Delta t/4}
  \right)
  \label{eq:g-defns}
  \end{equation}
}
in terms of the differences between the eigenstate masses ($\Delta m$)
and decay rates ($\Delta\Gamma$).

The time-independent complex parameters $a_{\pm}$ in
Equation~(\ref{eq:F12-defn}) can be written generally as
%
%
\ifthenelse{\boolean{istwocolumn}}{
%
  \begin{equation}
  a_{+}  =  \overline{\calA}_{\rm t} \: \calA_{\rm r} 
                   - \calA_{\rm t} \: \overline{\calA}_{\rm r}
  \ \ \ ,
  \end{equation}
  \begin{equation}
  \begin{split}
  a_{-}  =  -\sqrt{1-z^2}\,&\Big( \:
  \frac{q}{p} \: \overline{\calA}_{\rm t} \: \overline{\calA}_{\rm r} -
  \frac{p}{q} \: \calA_{\rm t} \: \calA_{\rm r}
  \Big)  \nonumber \\
   + \   z\,&\Big( \phantom{\frac{q}{p}} \!\!\!
  \overline{\calA}_{\rm t} \: \calA_{\rm r} +\calA_{\rm t} \:  \overline{\calA}_{\rm r}
  \Big) \,,
  \label{eq:a-defns}
  \end{split}
  \end{equation}
 %
 %
 %
}{
%
  \begin{equation}
  a_{+} = \overline{\calA}_{\rm t} \: \calA_{\rm r} 
                   - \calA_{\rm t} \: \overline{\calA}_{\rm r}
  \ \ \ , \ \ \
  a_{-} = -\sqrt{1-z^2}\,\left(
  \frac{q}{p} \overline{\calA}_{\rm t} \: \overline{\calA}_{\rm r} -
  \frac{p}{q} \calA_{\rm t} \: \calA_{\rm r}
  \right) +
  z\,\left(
  \overline{\calA}_{\rm t} \: \calA_{\rm r} +\calA_{\rm t} \:  \overline{\calA}_{\rm r}
  \right) \,,
  \label{eq:a-defns}
  \end{equation}
}
where $\calA_k$ ($\overline{\calA}_k$) is the \Bz (\Bzb) decay
amplitude to $f_k$.
The complex ratio $q/p$ parameterizes possible
$CP$ and $T$ violation ($|q/p|\ne 1$) in the time evolution of a neutral $B$ state,
while $z$, which is also complex, parametrizes possible $CPT$ and $CP$ violation
($z\neq0$) in the time evolution.
%
%
Note that exchanging the r and t subscripts changes the overall sign of $a_+$, $g_-$, and
$\Delta t$, leaving Eq.(\ref{eq:F12-defn}) unchanged, which is required since
the distinction between
the $B$ that is reconstructed and the $B$ that is used for flavor tagging
is arbitrary at this point.
Explicitly, we are using the conventions
\begin{equation}
\label{eq:qoverpdef}
\frac qp=-\sqrt{\frac{M_{12}^*-i\Gamma_{12}^*/2}{M_{12}-i\Gamma_{12}/2}}\,,
\end{equation}
where $M$ and $\Gamma$ are the hermitian matrices of the effective
Hamiltonian.
The eigenstates of the effective Hamiltonian are defined as
\begin{eqnarray}
  |B_L\rangle & = & p \: | \Bz\rangle + q \: | \Bzb \rangle \\
  |B_H\rangle & = & p \: | \Bz\rangle - q \: | \Bzb \rangle\,,
\end{eqnarray}
and $\Delta m = m_H - m_L$, which is positive by definition.
If $z=0$, as expected in the Standard Model, the two terms
$a_{\pm}g_{\pm}(\Delta t)$ in
Eq.(\ref{eq:F12-defn}) describe the cases where the surviving
meson undergoes a net oscillation ($-$) $\Bz\leftrightarrow\Bzb$
or not ($+$) between $t_{\rm t}$ and $t_{\rm r}$.
Combining Equations~(\ref{eq:F12-defn}--\ref{eq:a-defns}) we obtain
%
%
\ifthenelse{\boolean{istwocolumn}}{
%
  \begin{equation}
  \begin{split}
  F (\Delta t) = e^{-\Gamma |\Delta t|}\,\left[ \right.
  &\left. R\cosh(\Delta\Gamma\Delta t/2) + C \cos(\Delta m\Delta t) \right. \\
   + &\left.S '\sinh(\Delta\Gamma\Delta t/2) + S \sin(\Delta m\Delta t) \right]
  \end{split}
  \end{equation}
}{
%
  \begin{equation}
  F (\Delta t) =
  e^{-\Gamma |\Delta t|}\,\left[
  R\cosh(\Delta\Gamma\Delta t/2) + C \cos(\Delta m\Delta t) +
  S '\sinh(\Delta\Gamma\Delta t/2) + S \sin(\Delta m\Delta t)
  \right]
  \end{equation}
}
with coefficients which satisfy the constraint
$
{C }^2 + {S }^2 =
{R}^2 - {S '}^2\,,
$
and are given by
\begin{alignat}{4}
R &\equiv \frac{1}{2}\left(|a_{+}|^2 + |a_{-}|^2\right),
&\qquad\qquad&
S ' &\equiv -\text{Re}(a_{+}^\ast a_{-})\,, \notag \\
C   &\equiv \frac{1}{2}\left(|a_{+}|^2 - |a_{-}|^2\right),
&\qquad\qquad&
S   &\equiv  +\text{Im}(a_{+}^\ast a_{-})\,.
\label{eq:CS-defns}
\end{alignat}

In the following,
we assume CPT invariance so that $z=0$, and moreover we take
$\Delta \Gamma/\Gamma \ll 1$.
Thus the term $S'$ no longer enters and
$\cosh (\Delta\Gamma \Delta t/2)$ is replaced by unity.
The Standard Model predicts $|q/p|-1 = (2.5-6.5)\times10^{-4}$~\cite{Zoltan},
so we will assume $|q/p|\simeq 1$.
The resulting time dependence, when the tagged meson is a $\Bz$, is
\begin{equation}
F (\Delta t) =
e^{-\Gamma |\Delta t|}\,\left[
R + C \cos(\Delta m\Delta t) +
 S \sin(\Delta m\Delta t)
\right]\label{eq:simple_time_f}\,,
\end{equation}
and correspondingly when the tagged meson is a $\Bzb$
\begin{equation}
\overline{F} (\Delta t) =
e^{-\Gamma |\Delta t|}\,\left[
\overline R + \overline C \cos(\Delta m\Delta t) +
 \overline S \sin(\Delta m\Delta t)
\right] .
\label{eq:simple_time_fbar}
\end{equation}

\section{Characterization of the Tagging Amplitude}
\label{sec:assumptions}

The strength of the doubly-CKM-suppressed (DCS) decays can be expressed in
terms of the traditional parameter~\cite{Wolfenstein}
\begin{equation}
\lambda_f = \frac qp \frac {\overline{\calA}_f}{\calA_f}\,.
\end{equation}
This combination is independent of the choice of phases for the $\Bz$ and $\Bzb$ states.
Suppose $|f\rangle$ is a final state that is ostensibly the result of a
$\Bz$ decay.
For example, if $|f\rangle$ represents the tag $B$, a $K^+$ would indicate
that the tag $B$ decayed as a $\Bz$, assuming the dominant
$\overline{b}\rightarrow \overline{c}u\overline{d}$ transition occurred.
Then
\begin{equation}
\label{eq:LambdaDef}
\lambda_f = r_f e^{-2i\beta-i\gamma}e^{i\delta_f}\,,
\end{equation}
where $r_f$  is a real number of order 0.02 and $\delta_f$ is the
strong phase difference of the \Bzb decay relative to that of the \Bz decay,
assuming $\overline{b}\rightarrow \overline{c}u\overline{d}$ and
$b\rightarrow u \overline{c} d$ transitions for the \Bz and \Bzb decays respectively.
If, for this final state, there is only one mechanism contributing to the \Bz decay and to the
\Bzb decay, then for the $CP$ conjugate state $|\overline f\rangle$ we have
\begin{equation}
\lambda_f = \frac 1{r_f} e^{-2i\beta -i\gamma}e^{-i\delta_f}\,.
\end{equation}
We shall make the assumption of a single contributing amplitude except as
noted below.

Because the DCS amplitudes are only about 2\% of the
allowed amplitudes, in what follows we shall drop all terms that are quadratic or higher
in this suppression.
In practice we combine many final states $f$ in a single tagging category, $f\in T$.
For the tagging category we then have effective values of $r'$ and $\delta'$ defined by
\begin{equation}
r'e^{i\delta'}=\frac{\sum_{f\in T}\: \epsilon_f|\calA_f|^2 \: r_f e^{i\delta_f}}{  
\sum_{f\in T}\: \epsilon_f|\calA_f|^2}\,,
\label{eqn:reff}
\end{equation}
where $\epsilon_f$ is the relative tagging efficiency for the state $f$.
Notice that
\begin{equation}
|r'|\leq\frac{\sum_{f\in T}\: \epsilon_f|\calA_f|^2 |r_f|}{  
\sum_{f\in T}\: \epsilon_f|\calA_f|^2}\,,
\label{eqn:reffineq}
\end{equation}
so there is a tendency for contributions from different tagging states to
cancel, unless all contributions have nearly the same strong phase.
Equation~\ref{eqn:reff} holds only if terms of order $r_f^2$ can be ignored,
as we are assuming.

\section{Time-Dependent Asymmetry Coefficients}
\label{sec:asym-coefficients}

\begin{figure}[!t]
\begin{center}
\ifthenelse{\boolean{istwocolumn}}{
  \epsfig{figure=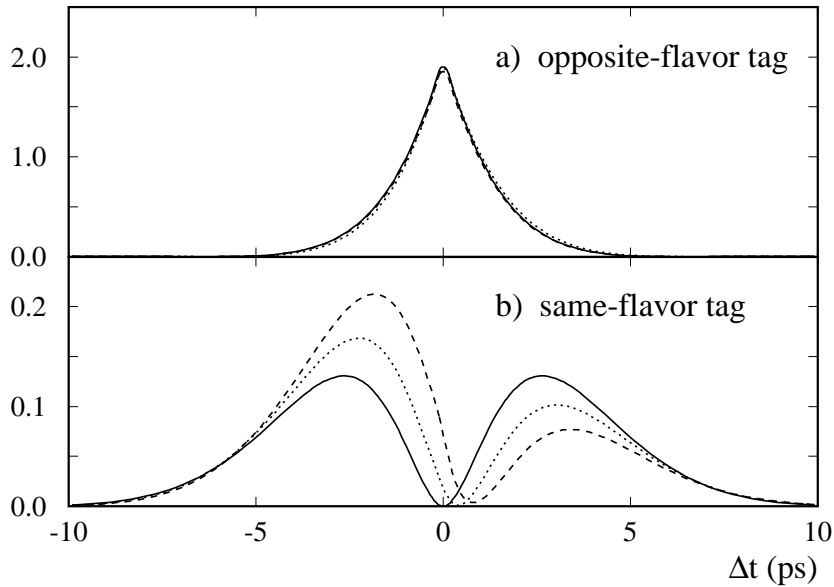,width=0.90\linewidth}
}{
  \epsfig{figure=fig2.eps,width=0.70\linewidth}
}
\caption{
Time-dependent decay distributions for the final state $D^{*-}\pi^+$, 
for a) a \Bzb tag, and b) a \Bz tag. $(2\beta+\gamma)$ is set to the value $1.86$. 
The situation with no doubly-CKM-suppressed
contribution on both the tag-side and reconstruction-side
is indicated with the solid line.
The dotted line has $r=0.1$ and $\delta=0$, but no tag-side interference.
The dashed line represents the example with $r=r'=0.1$, $\delta=0$, and $\delta'=\pi$.
In these examples, the $r$ and $r'$ values are $\times 5$ the expected values in order
to clearly illustrate the differences with respect to the case with $r=r'=0$.
}
\label{fig:mixingDt}
\end{center}
\end{figure}

In this section, we evaluate the coefficients
$R$($\overline{R}$),
$C$($\overline{C}$),
and $S$($\overline{S}$)
of Eqns.~\ref{eq:simple_time_f}(\ref{eq:simple_time_fbar}).
There are two specific cases that we will consider -- the ``mixing'' case,
where the reconstructed $B$ meson decays in an apparent flavor eigenstate
(e.g. $D^{*+} \pi^-$, normally assumed to originate from \Bzb decay),
and the ``$CP$'' case, where the reconstructed $B$ has decayed into a $CP$ eigenstate.
Dropping a common factor $\calA_{\rm t} \: \calA_{\rm r} \: (p/q)$, we can write
$a_+$ and $a_-$ in terms of the $\lambda$ parameters for the tag and reconstructed
$B$ mesons as
\begin{eqnarray}
a_+&=&\lambda_{\rm t}-\lambda_{\rm r}\nonumber\\
a_-&=&1-\lambda_{\rm t} \: \lambda_{\rm r}.
\end{eqnarray}
Quite generally then,
\begin{eqnarray}
\!\!\! |a_+|^2&\!=\!&|\lambda_{\rm t}|^2 -2 \: {\rm Re} \: \lambda_{\rm t} \: \lambda_{\rm r}^* 
     +|\lambda_{\rm r}|^2\nonumber\\
\!\!\! |a_-|^2&\!=\!&1 -2 \: {\rm Re}  \: \lambda_{\rm t} \: \lambda_{\rm r} 
     +|\lambda_{\rm t}|^2 \: |\lambda_{\rm r}|^2\nonumber\\
\!\!\! {\rm Im} \: a_+^*a_-&\!=\!&{\rm Im} \, \lambda_{\rm r} \: (1-|\lambda_{\rm t} |^2)
    \: - \: {\rm Im}\, \lambda_{\rm t} \: (1-|\lambda_{\rm r}|^2)\,.
\end{eqnarray}
Table~\ref{table:fourcases} gives the coefficients for the mixing case,
where for the reconstructed $B$ meson final state we have dropped the 
subscript $f$ from the amplitude ratio $r$ and from the strong phase difference $\delta$ 
in $\lambda_{\rm r}$, defined by Eq.(\ref{eq:LambdaDef}).
The only deviation from the familiar case with no
DCS
contributions, to first order in $r$ and $r'$, is the
presence of a small $S$($\overline{S}$) coefficient.
Figure~\ref{fig:mixingDt} shows an illustration of the time evolution for
when the flavor of the two $B$ mesons at the time of decay was opposite
(unmixed) or the same (mixed).
The nominal ($r=r'=0$) case is contrasted with an example of a non-zero
DCS
contribution in the reconstructed $B$ amplitude and
with an example of non-zero
DCS
contributions to both the tag
and reconstructed $B$ amplitudes.
The amplitude ratios $r$ and $r'$ have been enlarged by $\times5$ with respect
to the expected value (0.02) so that the DCS contributions are more clear.

\begin{figure}[!b]
\begin{center}
\ifthenelse{\boolean{istwocolumn}}{
  \epsfig{figure=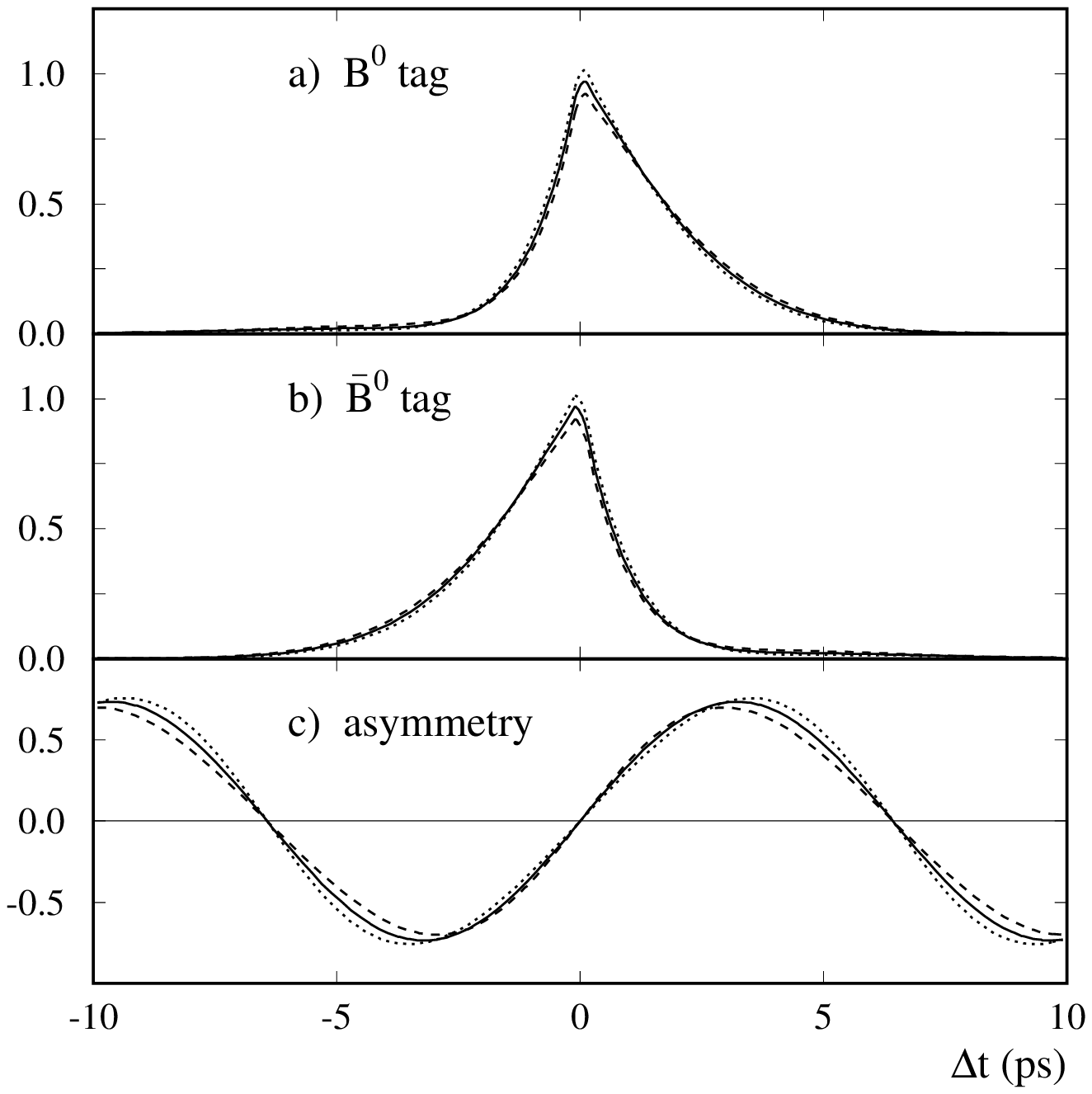,width=0.9\linewidth}
}{
  \epsfig{figure=fig3.eps,width=0.7\linewidth}
}
\caption{
Time-dependent decay distributions for the $\CP$ eigenstate $J/\psi K^0_S$, 
with a) a \Bz tag, and b) a \Bzb tag. We set $(2\beta+\gamma)$ to $1.86$. 
The situation with no tag-side interference is indicated with the solid line. 
The dotted line represents the case with $r'=0.1$ and $\delta'=0$,
and the dashed line has $r'=0.1$, and $\delta'=\pi$. 
It should be noted that, adding a non-zero DCS contribution, the slope and amplitude of the time-dependent asymmetry work in opposite directions.
In these examples, the $r'$ value is $\times 5$ the expected value in order
to clearly illustrate the differences with respect to the case with $r'=0$.
}
\label{fig:sin2bplot}
\end{center}
\end{figure}

Table~\ref{table:CP} gives the coefficients for the $CP$ case.
All three coefficients receive corrections linear in $r'$.
Figure~\ref{fig:sin2bplot} is an illustration of the the corrections
to the time evolution for \Bz and \Bzb tagged $CP$ events, also
with the
DCS
amplitude ratio $r'$ enlarged by $\times5$
to make the differences more visible.


\begingroup
\ifthenelse{\boolean{istwocolumn}}{ }{\squeezetable}
\begin{table*}\begin{center}
\begin{tabular}{ccccc}\hline\rule[-1em]{0pt}{3em} 
%
&[tag=$\Bz(K^+)$, rec=\Bz] \ \ & \ \ [tag=$\Bz(K^+)$,\ rec=\Bzb] \ \  &
 \ \ [tag=$\Bzb(K^-)$,\ rec=\Bz]  \ \ & \ \ [tag=$\Bzb(K^-)$,\ rec=\Bzb] \\
\hline\rule[-1em]{0pt}{3em} 
$\lambda_{\rm t}$ &$ r' e^{-2i\beta-i\gamma+i\delta'}$&$ r' e^{-2i\beta-i\gamma+i\delta'}$&$\frac 1{r'}e^{-2i\beta-i\gamma-i\delta'}$&$\frac 1{r'}e^{-2i\beta-i\gamma-i\delta'}$\\
\rule[-1em]{0pt}{3em}
$\lambda_{\rm r}$ &$ r e^{-2i\beta-i\gamma+i\delta}$&$\frac 1{r}e^{-2i\beta-i\gamma-i\delta}$&$ r e^{-2i\beta-i\gamma+i\delta}$&$\frac 1{r}e^{-2i\beta-i\gamma-i\delta}$\\
\rule[-1em]{0pt}{3em}
$|a_+|^2$&$0$&$\frac 1{r^2}$&$\frac 1{r'^2}$&$0$\\
\rule[-1em]{0pt}{3em}
$|a_-|^2$&$1$&$0$&$0$&$\frac 1{+r^2r'^2}$\\
\rule[-1em]{0pt}{3em}
Im $a_+^*a_-$ & $-r\sin(2\beta+\gamma-\delta)$& $-\frac{r'}{r^2}\sin(2\beta+\gamma-\delta')$&$\frac r{r'^2}\sin(2\beta+\gamma-\delta)$&$-\frac 1{r^2r'}\sin(2\beta+\gamma+\delta')$\\ 
              & $+r'\sin(2\beta+\gamma-\delta')$&$-\frac 1{r}\sin(2\beta+\gamma+\delta)$&$+\frac 1{r'}\sin(2\beta+\gamma+\delta')$&$+\frac 1{rr'^2}\sin(2\beta+\gamma+\delta)$ \\
              \rule[-1em]{0pt}{3em}
$R$ &$1$&$1$&$1$&$1$\\ \rule[-1em]{0pt}{3em}
$C$ &$-1$&$1$&$1$&$-1$\\ \rule[-1em]{0pt}{3em}
$S$ & $-2r\sin(2\beta+\gamma-\delta)$& $-2r'\sin(2\beta+\gamma-\delta')$&$ 2r\sin(2\beta+\gamma-\delta)$&$-2r'\sin(2\beta+\gamma+\delta')$\\
              & $+2r'\sin(2\beta+\gamma-\delta')$&$-2r\sin(2\beta+\gamma+\delta)$&$+2r'\sin(2\beta+\gamma+\delta')$&$+2r\sin(2\beta+\gamma+\delta)$
\\ \\
\hline\hline
\end{tabular}
\caption[a]{Contributions to the time dependence of tagged decays when the
reconstructed decay is an apparent flavor eigenstate, with
doubly-CKM-suppressed decays considered only to first order.  The
dependences proportional to $1$ and to $\cos\Delta m\Delta t$ are
unaffected.  A small $\sin\Delta m\Delta t$ term is induced.
Appropriate factors of $r$ and $r'$ have been removed to scale $R$ to
unity.}
\label{table:fourcases}
\end{center}
\end{table*}
\endgroup


\begingroup
\begin{table*}\begin{center}
\begin{tabular}{ccc}\hline\rule[-1em]{0pt}{3em} 
&[tag=$\Bz(K^+)$,\ rec=$B_{CP}$] & [tag=$\Bzb(K^-)$,\ rec=$B_{CP}$]  \\
\hline\rule[-1em]{0pt}{3em}
$\lambda_{\rm t}$ &$ r' e^{-2i\beta-i\gamma+i\delta'}$&$\frac 1{r'}e^{-2i\beta-i\gamma-i\delta'}$\\ 
\rule[-1em]{0pt}{3em}
$\lambda_{\rm r}$ &$\lambda_{CP}$&$\lambda_{CP}$\\
\rule[-1em]{0pt}{3em}
$|a_+|^2$&$|\lambda_{CP}|^2-2 \: {\rm Re}\:  r'e^{-2i\beta-i\gamma+i\delta'}\lambda_{CP}^*$&$\frac 1{r'^2}-2{1\over r'}\: {\rm Re}\:  \lambda_{CP}e^{2i\beta+i\gamma+i\delta'}$\\
\rule[-1em]{0pt}{3em}
$|a_-|^2$&$1-2 \: {\rm Re}\:  r'e^{-2i\beta-i\gamma+i\delta'}\lambda_{CP}$&$\frac {|\lambda_{CP}|^2}{r'^2}-2{1\over r'}\:  {\rm Re}\:  \lambda_{CP}e^{-2i\beta-i\gamma-i\delta'}$\\
\rule[-1em]{0pt}{3em}
Im $a_+^*a_-$ & \ \ \ $ {\rm Im} \: \lambda_{CP}+r'(1-|\lambda_{CP}^2|)\sin(2\beta+\gamma-\delta')$ \ \ \ & \ \ \ $-\frac 1{r'^2}  \:{\rm Im} \: \lambda_{CP}+\frac 1{r'}(1-|\lambda_{CP}|^2)\sin(2\beta+\gamma+\delta')$ \ \ \ \\
\rule[-1em]{0pt}{3em}
$R$ &${1+|\lambda_{CP}^2|\over 2}-2r' \: {\rm Re}\:  \lambda_{CP}\cos(2\beta+\gamma-\delta')$&${1+|\lambda_{CP}^2|\over 2}-2r' \: {\rm Re}\:  \lambda_{CP}\cos(2\beta+\gamma+\delta')$\\
\rule[-1em]{0pt}{3em}
$C$ &${|\lambda_{CP}^2|-1\over 2}+2r'  \:{\rm Im} \: \lambda_{CP}\sin(2\beta+\gamma-\delta')$&${1-|\lambda_{CP}^2|\over 2}+2r'  \:{\rm Im} \: \lambda_{CP}\sin(2\beta+\gamma+\delta')$\\
\rule[-1em]{0pt}{3em}
$S$ & $  \:{\rm Im} \: \lambda_{CP}+r'(1-|\lambda_{CP}^2|)\sin(2\beta+\gamma-\delta')$&$-  \:{\rm Im} \: \lambda_{CP}+r'(1-|\lambda_{CP}^2|)\sin(2\beta+\gamma+\delta')$\\
        \\ \hline\hline
\end{tabular}
\caption{Contributions to the time dependence of tagged decays when the 
reconstructed decay is a CP eigenstate, with doubly-CKM-suppressed decays considered only to first order.  Appropriate factors of $r$ and $r'$ have been removed to scale
$R$ to unity in the limit in which the doubly-CKM-suppressed decays vanish.}
\label{table:CP}
\end{center}
\end{table*}
\endgroup


\section{Completely Inclusive Tagging Categories}
\label{sec:perfect-inclusive}

We can relate the effective $r'$ and $\delta'$ to the $2\times2$ matrix $\Gamma$ that
generalizes the decay rate for the \BzBzb system.
Let $T$ be the class of states $DX$, where $X$ represents non-charmed hadrons.
Neglecting the relative tagging efficiency $\epsilon_f$ for the moment, we have
%
%
\ifthenelse{\boolean{istwocolumn}}{
%
%
   \begin{eqnarray}
   \sum_{f\in T}\frac qp \calA_f^*{\overline \calA}_f & = & \sum_{f\in T} |\calA_f|^2\lambda_f \nonumber \\
    & = &\sum_{f\in T} \langle \Bz|{\cal H}|f\rangle\langle f|{\cal H}|\Bz\rangle r_f e^{-2i\beta-i\gamma+i\delta_f } \nonumber \\
   \label{eq:PerfectInclusiveSum}
    & = &\Gamma_{DX}\,  r'e^{-2i\beta-i\gamma+i\delta '}\,,
   \end{eqnarray}
}{
%
   \begin{equation}
   \label{eq:PerfectInclusiveSum}
   \sum_{f\in T}\frac qp \calA_f^*{\overline \calA}_f=\sum_{f\in T}
   |\calA_f|^2\lambda_f=\sum_{f\in T}
   \langle \Bz|{\cal H}|f\rangle\langle f|{\cal H}|\Bz\rangle r_f e^{-2i\beta-i\gamma+i\delta_f }=\Gamma_{DX}\,  r'e^{-2i\beta-i\gamma+i\delta '}\,,
   \end{equation}
}
where $\Gamma_{DX}$ is, up to a trivial normalization, the partial width of
$\Bz$ into the class of states of the form $DX$.
On the other hand, we can write
   \begin{equation}
   \sum_{F\in T}\frac qp \calA_f^*{\overline \calA}_f=\sum_{F\in T}\frac qp\langle \Bz|{\cal H}|f\rangle\langle f|{\cal H}|\Bzb\rangle
   =\frac qp\Gamma_{DX\, 12}\,,
   \end{equation}
where $\Gamma_{DX\, 12}$ is the contribution of states of the form  $DX$ to
the off-diagonal part of the $\Gamma$ matrix.
 So 
\begin{equation}
 r'e^{-2i\beta-i\gamma+i\delta '}=\frac qp\Gamma_{DX\ 12}/\Gamma_{DX}\,.
\end{equation}
If tagging does not capture every state, we can think of $\Gamma_{12}$ and
$\Gamma$ as effective quantities, limited by the partial sum over states.
However, if that sum were complete, then $\delta '$ would vanish.  To see this,
imagine using as a basis of states $f_S$ not the physical states that are observed
but instead a basis of states that are eigenstates of the S matrix, that is,
a basis of states that each scatter into themselves. Because we are summing over
all states in a collection connected by strong interactions, there is such a
basis. Then the final state
interaction phases associated with $\calA_{f_S}$ and ${\overline\calA}_{f_S}$
would both be $e^{i\delta_{f_S}}$.  These would cancel in $\calA_{f_S}^*
{\overline\calA}_{f_S}$.
In general, because tagging is incomplete, we
cannot assume that $\delta '$ vanishes.
In reality, the relative tagging efficiency $\epsilon_f$, set to one in
Eq.(\ref{eq:PerfectInclusiveSum}), is not the same for all of the states in $T$, so
the tagging category representing the class of states $DX$ is not completely inclusive.

\section{Estimated Size of Doubly-CKM Suppressed Amplitude}
\label{sec:rprime-and-deltaprime}

In the Introduction, we gave an estimate for the size of the
DCS
amplitude ($r$), relative to the favored
amplitude, to be approximately 0.02, which is simply the
ratio of the CKM elements involved,
$|(V_{ub}^* V_{cd}^{\vphantom{*}}) / (V_{cb}^{\vphantom{*}} V_{ud}^*)|$.
Here, we discuss the uncertainty of this estimate as
well as what can be assumed, if anything, about the strong phase
difference ($\delta$) between the
DCS
and favored amplitudes.

We use measured charm branching fractions
as a test of our simple amplitude ratio estimate.
The charm decay $D^0 \rightarrow K^+\pi^-$ is doubly-CKM suppressed
relative to the favored $D^0 \rightarrow K^-\pi^+$ decay.
The amplitude ratio prediction, based solely on the
CKM elements, gives
$r\approx |(V_{cd}^* V_{us}^{\vphantom{*}})/(V_{cs}^* V_{ud}^{\vphantom{*}})| \approx 0.048$ .
The experimental value from the branching fractions~\cite{PDG} is $0.062\pm0.005$,
which is within 25\% of the ratio of CKM elements.
For the singly-CKM suppressed charm decays $D^0 \rightarrow K^+K^-$ and
$D^0 \rightarrow \pi^+\pi^-$
we would estimate amplitude ratios relative to the allowed amplitude of
$|V_{us}/V_{ud}| \approx |V_{cd}/V_{cs}| \approx 0.23$,
while the branching ratios give
$0.329\pm0.007$ and $0.194\pm0.005$
for $K^+K^-$ and $\pi^+\pi^-$,
respectively.

The decay $B^0\rightarrow D^+ \pi^-$ is doubly-CKM suppressed, but
this branching fraction has not been measured.
We can estimate its branching fraction from the related decay mode
$B^0\rightarrow D^+_s \pi^-$, which has been
observed recently~\cite{Dspi}, and has a branching fraction of
$(3.2 \pm 0.9 \pm 1.0)\times 10^{-5}$.
The amplitude ratio for $B^0\rightarrow D^+ \pi^-$, relative to
$B^0\rightarrow D^- \pi^+$ is estimated to be
\[
   r_{D\pi} \approx \sqrt{\frac{{\cal B}(B^0\rightarrow D^+_s\pi^-)}
                               {{\cal B}(B^0\rightarrow D^-\pi^+)}}
                 \left| \frac{V_{cd}}{V_{cs}} \right|
                 \frac{f_D}{f_{D_s}} \approx 0.021 \pm 0.005 \ ,
\]
where we have used $f_D/f_{D_s} = 1.11 \pm 0.01 \pm 0.01$ (from~\cite{lattice})
to approximate $SU(3)$ breaking effects.
This is in good agreement with the naive estimate of 0.020, albeit with
a large uncertainty.

There are some theoretical arguments for expecting the strong
phase difference $\delta$ to be small~\cite{smallStrongPhase}, but
we know of at least one case where a non-trivial strong phase has
been observed in $B$ decay.
The strong phase difference between the longitudinal and parallel
polarization amplitudes of the transversity basis
in $B\rightarrow J/\psi K^*(892)$ has been
measured~\cite{babarJpsikstar} to be $2.50 \pm 0.22$, which is about
$3\sigma$ from $\pi$, in contradiction with the factorization prediction of 0 or $\pi$.
The size of the effective amplitude ratio ($r'$), given by
Equation~\ref{eqn:reff}, depends on the $\delta$ values of
the final states included in the tagging category.
As Eq.~\ref{eqn:reffineq} shows, varying $\delta$ values
between the states will tend to reduce $r'$.

Given the $\approx 50\%$ uncertainty on the
DCS
amplitude ratio
$r$ for individual final states and the general lack of knowledge
concerning strong phase differences, we conclude that the most conservative
assumptions regarding the effective parameters $r'$ and $\delta'$
would be to allow $r'$ values from 0 (full cancellation in the sum)
up to 0.04 (no cancellation with some enhancement over our 0.02 estimate)
and to allow any value of $\delta'$.

\section{Uncertainties in Fitted Asymmetries}
\label{sec:asym-uncertainty}

In this section, we will discuss the uncertainties
due to tag-side interference
on some common time-dependent asymmetries.
In addition to the assumptions that we have already made
(i.e. $z=0$, $\Delta \Gamma/\Gamma = 0$, and $|q/p|=1$), one usually
assumes that the tag-side amplitude is dominated by a
single contribution, or $r'=0$.
The time dependent coefficients in Eqns.~\ref{eq:simple_time_f}
and~\ref{eq:simple_time_fbar} simplify considerably with this
assumption.
For the case where the reconstructed $B$ is a $CP$ eigenstate,
we have
\begin{equation}
\label{eqn:CPAssumptions}
  R_{CP} = \overline{R}_{CP} \ \ \ , \ \ \
  C_{CP} = - \overline{C}_{CP} \ \ \ , \ \ \
  S_{CP} = - \overline{S}_{CP}\,,
\end{equation}
which can be seen from Table~\ref{table:CP} with $r'$ set to zero.
For the case where the reconstructed $B$ is in an apparent flavor
eigenstate, the coefficients in Table~\ref{table:fourcases} with
$r'=0$ give
\begin{equation}
\label{eqn:MixAssumptions}
  R_{\rm mix} = R_{\rm unmix} \  \ , \  \
  C_{\rm mix} = - C_{\rm unmix} \  \ , \  \
  S_{\rm mix} = S_{\rm unmix} = 0\,,
\end{equation}
where the ``mix'' (``unmix'') subscript refers to the case where the tag
and reconstructed $B$ mesons were the same (opposite) flavor
at the time of decay.
In the rest of this Section, we will evaluate the bias on the fitted
coefficients when fitting the data with the assumptions in Eqns.~\ref{eqn:CPAssumptions} 
or~\ref{eqn:MixAssumptions} of nonzero tag-side interference.

In the relations above, the $R$ coefficients are independent of
the final state configuration, so they are usually absorbed into the
$C$ and $S$ coefficients by fitting for ${\cal C} \equiv (C/R)$ and ${\cal S} \equiv (S/R)$.
A fairly reliable estimate of the fitted ${\cal C}$ coefficient is simply the
asymmetry at $\Delta t = 0$.
This would be
\begin{equation}
\label{eqn:estC}
  {\cal C}_{\rm fit} \approx \frac{ C + R - \overline{C} - \overline{R}}
                                  { C + R + \overline{C} + \overline{R}}
\end{equation}
for a $CP$ asymmetry, or
\begin{equation}
\label{eqn:estCmix}
  {\cal C}_{\rm fit} \approx \frac{ C_{\rm unmix} + R_{\rm unmix} - C_{\rm mix} - R_{\rm mix}}
                                  { C_{\rm unmix} + R_{\rm unmix} + C_{\rm mix} + R_{\rm mix}}
\end{equation}
for a mixing asymmetry.
A similar, but slightly less reliable, estimate for the fitted ${\cal S}$ coefficient
in a $CP$ asymmetry is simply the flavor-averaged ${\cal S}$ coefficient,
or
\begin{equation}
\label{eqn:estS}
  {\cal S}_{\rm fit} \approx \frac{1}{2}\left( \frac{S}{R} - \frac{\overline{S}}{\overline{R}} \right).
\end{equation}

Precise estimates can be derived using a simple maximum likelihood technique, where
the likelihood to be maximized with respect to ${\cal C}_{\rm fit}$ and ${\cal S}_{\rm fit}$ is
%
%
%
\ifthenelse{\boolean{istwocolumn}}{
%
  \begin{equation}
  \label{eqn:logl}
  \begin{split}
   {\cal L}  =  N \int_{-\infty}^{\infty} \!\!\!\!\! d\Delta t \ e^{-\Gamma |\Delta t|}
     [ &F(\Delta t) \ \ln F_{\rm fit}(\Delta t) \nonumber  \\
         + \ &\left.\!\! \overline{F}(\Delta t) \ \ln \overline{F}_{\rm fit}(\Delta t) \right],
   \end{split}
  \end{equation}
}{
  \begin{equation}
  \label{eqn:logl}
   {\cal L} = N \int_{-\infty}^{\infty} d\Delta t \ e^{-\Gamma |\Delta t|}
     \left[ F(\Delta t) \ \ln F_{\rm fit}(\Delta t)
          + \overline{F}(\Delta t) \ \ln \overline{F}_{\rm fit}(\Delta t)
     \right],
  \end{equation}
}
with $F_{\rm fit}$ and $\overline{F}_{\rm fit}$ evaluated using the assumptions
in Eq.~\ref{eqn:CPAssumptions}.
We confirmed that Eqns.~\ref{eqn:estC} and~\ref{eqn:estS} give reasonable estimates
of ${\cal C}_{\rm fit}$ and ${\cal S}_{\rm fit}$ with unbinned maximum likelihood fits of simulated
data samples.

\subsection{Mistag calibration with flavor oscillation amplitude}
\label{sec:mistag}

As was mentioned above, the sign of the tagging kaon charge does not always
give the correct flavor tag.
For example, CKM-suppressed $D$ decays, such as
$D^+ \rightarrow K^+ \overline{K}$$^0$,
can produce wrong-sign kaons.
Pions, incorrectly identified as kaons, can also produce
wrong-sign kaons.
The amplitude of any measured asymmetry using kaon tags will be reduced
by a factor of $(1-2\omega)$, sometimes called the dilution
factor, where $\omega$ is the fraction of tagging
kaons that have the wrong sign (mistag fraction).
The mistag fraction $\omega$ is usually measured from the amplitude
of time-dependent flavor oscillations in a sample of reconstructed $B^0$
decays to flavor-specific final states~\cite{s2b-mix-PRD}.
The measured value of ${\cal C}$ will be a direct measurement
of $(1-2\omega)$, which can then be used to translate measured $CP$
asymmetry coefficients.

To first order in $r$ and $r'$, the $R$ and $C$ coefficients are
the expected ones, as can be seen in Table~\ref{table:fourcases}.
The only effect is in the $S$ coefficient, which is usually
assumed to be zero in the analysis of mixing data.
This means that the measured mistag fractions will be unaffected by
DCS
amplitude contributions, either on the
tag side or the reconstructed side, since our estimator for ${\cal C}_{\rm fit}$ only
depends on the $R$ and $C$ coefficients.
Contrary to what one may guess, the corrections due to
DCS
amplitude contributions are not simply absorbed
into the mistag fractions.

Using Monte Carlo pseudo-experiments, we also find that $\Delta m_{d}$ is unaffected
to the level of $0.001$ ps$^{-1}$ if allowed to float in the fit.

\subsection{Fully reconstructed CP eigenstates}
\label{sec:sin2beta}

The size of $CP$ asymmetries in $B$ decays to $CP$ eigenstates are in general of order one in the
Standard Model.
For example, $CP$ asymmetry in $B \rightarrow J/\psi K^0_S$ (and related
charmonium modes) has been measured to be ${\cal S}_{\rm fit} = 0.735\pm 0.055$ ~\cite{babarS2B,belleS2B}.
Any deviations due to tag-side interference 
$(\approx 0.02)$ will be comparatively small (see Fig.~\ref{fig:sin2bplot}), and can be treated 
as perturbations on the usual measurements.

In what follows, the nominal values for the fitted $CP$ asymmetry coefficients without
any tag-side interference from doubly-CKM suppressed decays are defined as
\begin{eqnarray}
  {\cal C}_0  & = & \frac{|\lambda_{CP}|^2 - 1}
                         {|\lambda_{CP}|^2 + 1} \\
  {\cal S}_0  & = & \frac{ 2 {\rm Im} \lambda_{CP}}
                         {|\lambda_{CP}|^2 + 1}.
\end{eqnarray}
The expected fitted coefficients,
when the fit is performed with the assumptions in Eq.~\ref{eqn:CPAssumptions},
can be found by inserting the $R$, $C$, and $S$ values from Table~\ref{table:CP} into
Eqns.~\ref{eqn:estC} and~\ref{eqn:estS}.
Working to first order in $r'$, we find
%
%
%
\ifthenelse{\boolean{istwocolumn}}{
%
%
  \begin{equation}
  \begin{split}
    \label{eqn:cpCfit}
    {\cal C}_{\rm fit}  =   {\cal C}_0 \left[ 1 \right.
        + \ &  \left.\! 2 r' \cos \delta' \left\{ {\cal G} \cos(2\beta+\gamma) -
              {\cal S}_0 \sin(2\beta+\gamma) \right\}\right]  \\
        - \ & 2 r' \sin \delta' \left\{ {\cal S}_0 \cos(2\beta+\gamma) + {\cal G} \sin(2\beta+\gamma) \right\} \\
  \end{split}
  \end{equation}
  \vspace{-0.5cm}
  \begin{multline}
  \label{eqn:cpSfit}
  \!\!\!\!\!\!\!  \makebox[\columnwidth][l]{$  {\cal S}_{\rm fit}   =   {\cal S}_0 \left[ 1  +   2 r' \cos \delta' {\cal G} \cos(2\beta+\gamma) \right]$}  \\
   \makebox[0.68\columnwidth][l]{$ + 2 r' \sin \delta' {\cal C}_0 \cos(2\beta+\gamma)\,, $}
  \end{multline}
%
%
%
}{
%
%
\begin{eqnarray}
  \label{eqn:cpCfit}
  {\cal C}_{\rm fit}   =   {\cal C}_0 \left[ 1 \right.
    &  + & \left. 2 r' \cos \delta' \left\{ {\cal G} \cos(2\beta+\gamma) -
            {\cal S}_0 \sin(2\beta+\gamma) \right\}\right] \nonumber \\
  & - & 2 r' \sin \delta' \left\{ {\cal S}_0 \cos(2\beta+\gamma) + {\cal G} \sin(2\beta+\gamma) \right\} \\
& & \nonumber \\
\label{eqn:cpSfit}
  {\cal S}_{\rm fit}   =   {\cal S}_0 \left[ 1 \right.
    & + & \left. 2 r' \cos \delta' {\cal G} \cos(2\beta+\gamma) \right]
    + 2 r' \sin \delta' {\cal C}_0 \cos(2\beta+\gamma)\,,
\end{eqnarray}
}
where ${\cal G} \equiv 2 {\rm Re} \lambda_{CP} / ( |\lambda_{CP}|^2 + 1 )$.
Note that, with respect to the nominal values, there are both multiplicative 
and additive corrections which are proportional to $\cos \delta'$ and $\sin \delta'$ respectively.
In the limit of a vanishing effective tag-side strong phase difference
($\delta' \rightarrow 0$), only the multiplicative corrections remain.

For $B^0 \rightarrow J/\psi K^0_S$, the dominant tree and penguin
amplitude contributions share the same weak phase.
The highly suppressed $u$-quark penguin, which has a different relative
weak phase, is typically ignored, giving the
Standard Model prediction of $\lambda_{J/\psi K^0_S} = - e^{-i2\beta}$.
Inserting this into Eqns.~\ref{eqn:cpCfit} and~\ref{eqn:cpSfit} gives
\ifthenelse{\boolean{istwocolumn}}{
\begin{align}
  \label{eqn:sin2betaCfit}
   {\cal C}_{\rm fit}[J/\psi K^0_S] &=  -2 r' \sin \gamma \sin \delta' \\
  \label{eqn:sin2betaSfit}
   {\cal S}_{\rm fit}[J/\psi K^0_S] &=  {\cal S}_0 \left[ 1
         - 2 r' \cos \delta' \left\{ \
               \cos 2\beta \cos(2\beta + \gamma) \nonumber  \right. \right.\\
   &  \ \ \ \  + {\cal K} \  \left. \left. \sin 2 \beta \sin(2\beta+\gamma) \right\}
               \right],
\end{align}
}{
\begin{eqnarray}
  \label{eqn:sin2betaCfit}
   {\cal C}_{\rm fit}[J/\psi K^0_S] & = & -2 r' \sin \gamma \sin \delta' \\
  \label{eqn:sin2betaSfit}
   {\cal S}_{\rm fit}[J/\psi K^0_S] & = & {\cal S}_0 \left[ 1
         - 2 r' \cos \delta' \left\{
               \cos 2\beta \cos(2\beta + \gamma)
               + {\cal K} \sin 2 \beta \sin(2\beta+\gamma) \right\}
               \right],
\end{eqnarray}
}
with ${\cal C}_0 = 0$ and  ${\cal S}_0 = \sin 2 \beta$.
The last term in Eq.~\ref{eqn:sin2betaSfit} proportional to ${\cal K}$ is a correction
to the simple estimate given by Eq.(\ref{eqn:cpSfit}).
The correction ${\cal K}$ was derived from the more precise likelihood analysis
given by Eq.(\ref{eqn:logl}).
The value of ${\cal K}$ is between 0.10 and 0.35, depending on the value of $\sin 2\beta$.
If we assume $\sin 2\beta = 0.74$ and allow $\gamma$ to be in the
range [$45^\circ$,$90^\circ$], then ${\cal K}=0.28$ and
the magnitude of the deviation of ${\cal S}_{\rm fit}$ away from
the nominal value ${\cal S}_0$ is $<0.7 \: r$.
The size of the deviation of ${\cal C}_{\rm fit}[J/\psi K^0_S]$
could be as large as $2 \: r'$.
These corrections to ${\cal S}_{\rm fit} = {\cal S}_0$
and ${\cal C}_{\rm fit} = 0$ could be as large or larger than Standard
Model corrections~\cite{ligeti}.

The uncertainty estimates in the previous paragraph apply to a measurement
that only uses kaon tags.
In practice, all useful sources of flavor information from the tag
side $B$ are employed in order to maximize the sensitivity of the measurement.
The statistical error on the measured asymmetry scales as $1/\sqrt{\sum_i Q_i}$,
where each flavor tagging category contributes $Q_i = \epsilon_i (1-2\omega_i)^2$
and $\epsilon_i$ is efficiency for category $i$.
Lepton flavor tags do not have the problem of a suppressed amplitude contribution
with a different weak phase, so we assume that $r'=0$ for lepton tags.
If a measurement uses both lepton and non-lepton tags, the magnitude of
the tag-side interference uncertainty will be scaled down by a factor of
$Q_{\rm non-lep}/(Q_{\rm lep} + Q_{\rm non-lep})$.
For example, the BaBar flavor tagging algorithm\cite{babarS2B} has roughly
$Q_{\rm lep}\approx 0.1$ and $Q_{\rm non-lep}\approx 0.2$.
This gives a reduction of the tag-side interference uncertainty of about
a factor of 2/3.

The $CP$ asymmetry for $B \rightarrow \pi^+ \pi^-$ is more complex.
This decay has both tree and penguin amplitude contributions which are comparable in
magnitude, have different weak phases, and have an experimentally unknown relative strong
phase difference.
Equations~\ref{eqn:cpCfit} and~\ref{eqn:cpSfit} do not become more transparent after inserting
the value for $\lambda_{\pi\pi}$ given below
\begin{equation}
   \lambda_{\pi\pi} = e^{-2i(\beta+\gamma)} \left( \frac{1 + |P/T| e^{i\delta} e^{i\gamma} }
                                                       {1 + |P/T| e^{i\delta} e^{-i\gamma} } \right)\,,
\end{equation}
where the $t$-quark penguin has been absorbed into the tree and penguin amplitudes
using unitarity of the CKM matrix, as in~\cite{GRprd65}.
Clearly, both the reconstructed and tag $B$ amplitudes now depend on $\gamma$, so care
must be taken in evaluating the tag-side interference uncertainty, which in general
can be as large as $2 \: r'$ for either the multiplicative or additive terms in
Eqns.~\ref{eqn:cpCfit} and~\ref{eqn:cpSfit}.
%

\section{Measurement of {\boldmath $\sin(2\beta+\gamma)$} with {\boldmath $D^{(*)}\pi$}}
\label{sec:sin2bpg}

One technique for measuring or constraining $\gamma$ is to perform a
time-dependent analysis of a decay mode that is known to have a non-zero DCS contribution,
such as $D^{*+}\pi^-$~\cite{Dstpi-2bpg}.
The time-dependent asymmetry coefficients are those given in Table~\ref{table:fourcases}.
In the usual case, tag-side interference is ignored ($r'=0$) and the amplitude
of the $\sin \Delta m \Delta t$ term is $2 r \sin(2\beta+\gamma \pm \delta)$,
where $r$ is the ratio of the DCS to CKM-favored amplitude contributions for the
reconstructed, or non-flavor-tag, $B$ and $\delta$ is the strong phase difference
between the two amplitudes.
Measuring $r$ and $\sin(2\beta+\gamma \pm \delta)$ simultaneously is very challenging,
so it is likely that $r$ will have to be constrained from other measurements~\cite{Dspi}.

\begingroup
\begin{table*}[ht!]
\begin{center}
\begin{tabular}{cllrl}
\hline
\rule[-1em]{0pt}{2em}
   Symbol & \multicolumn{1}{c}{Reco}  &  \multicolumn{1}{c}{Tag}
      &  \multicolumn{2}{c}{$\sin(\Delta m \Delta t)$ coefficient} \\
   \hline\hline
   \rule[-1em]{0pt}{2em}
     $S1$  & \ \ \ $B^0$ ($D^{*-}\pi^+$) \ \ \  &   \ \ \ $B^0$  ($K^+$) \ \ \
           & $-2 \: r \sin(2\beta+\gamma-\delta) \! $&$ + \ 2 \: r' \sin(2\beta+\gamma-\delta')$ \\
   \rule[-1em]{0pt}{2em}
     $S2$  & \ \ \ $B^0$ ($D^{*-}\pi^+$) \ \ \  &   \ \ \ $\Bzb$ ($K^-$) \ \ \
           & $ 2 \: r \sin(2\beta+\gamma-\delta) \! $&$ + \ 2 \: r' \sin(2\beta+\gamma+\delta')$ \\
   \rule[-1em]{0pt}{2em}
     $S3$  & \ \ \ $\Bzb$ ($D^{*+}\pi^-$) \ \ \ &   \ \ \ $B^0$  ($K^+$) \ \ \
           & $-2 \: r \sin(2\beta+\gamma+\delta) \! $&$ - \ 2 \: r' \sin(2\beta+\gamma-\delta')$ \\
   \rule[-1em]{0pt}{2em}
     $S4$  & \ \ \ $\Bzb$ ($D^{*+}\pi^-$) \ \ \ &   \ \ \ $\Bzb$ ($K^-$) \ \ \
           & $ 2 \: r \sin(2\beta+\gamma+\delta) \! $&$ - \ 2 \: r' \sin(2\beta+\gamma+\delta')$ \\
   \hline
%
%
%
%
%
\end{tabular}
\caption{The 4 coefficients of the $\sin \Delta m \Delta t $ term in the
  time-dependence of $D^*\pi$.  The 2nd and 3rd columns give the interpretation
  of the observed final state (given in parentheses) in terms of the dominant amplitude.}
\label{tab:2bpg-S}
\end{center}
\end{table*}
\endgroup

Since both $r$ and $r'$ are expected to be of the same order ($\approx 0.02$),
it is clear that tag-side DCS interference can not be treated as a perturbation
on the usual case. 
This effect is illustrated in Fig.~\ref{fig:mixingDt}. 
The time dependent analysis should be performed in a way that is general enough 
to accommodate $r' \approx r$ and any value of $\delta'$.

Table~\ref{tab:2bpg-S} gives the $\sin \Delta m \Delta t$ coefficients,
taken from Table~\ref{table:fourcases}, for the
4 combinations of reconstructed and flavor tag $B$ final states, where we
have neglected $r^2$, $rr'$, and $r'^2$ contributions.
It is useful to rewrite the relations for the $S$ coefficients in the following
way
\begin{eqnarray}
\label{eq:abcpars-s1}
   S1 & =&  -a + b + c \\
   S2 & =&  +a + b - c \\
   S3 & =&  -a - b - c \\
\label{eq:abcpars-s4}
   S4 & =&  +a - b + c\,,
\end{eqnarray} 
where the 3 variables to be determined in the time-dependent analysis are
\begin{eqnarray}
  \label{eqn:a}
   a & \equiv & 2 \: r \sin(2\beta+\gamma) \cos\delta \\
  \label{eqn:b}
   b & \equiv & 2 \: r' \sin(2\beta+\gamma) \cos\delta' \\
  \label{eqn:c}
   c & \equiv & 2 \cos(2\beta+\gamma) \left( r\sin\delta - r'\sin\delta'\right)\,.
\end{eqnarray}
This parameterization makes no assumptions about the magnitude of
$r'$ or $\delta'$, and is attractive for several reasons.
First, $a$ does not depend at all on the tag-side parameters $r'$ and $\delta'$.
In the case where $\delta=0$, which is favored by some~\cite{smallStrongPhase},
$a$ is exactly what one wants to know ($\sin(2\beta+\gamma)$).
Secondly, this parameterization cleanly separates the flavor-tag symmetric
and antisymmetric components; the $a$ and $c$ coefficients are diluted by
a factor of $(1-2\omega)$, while the $b$ coefficient is not, since it has
the same sign for tag-side \Bz and tag-side \Bzb events.
The minimum number of independent parameters
in which the $S$ coefficients can be written is three.
We recommend using the $a$, $b$, and $c$ coefficients as the experimental
parameters to be determined in the time-dependent asymmetry analysis.


The set of kaon tagging final states that yields correct tags is in
general quite different from the set of final states that yields incorrect
tags.
This means that within a tagging category, the effective $r'$ and
$\delta'$ values for correct tags are different from those for
incorrect tags.
In the sum over correct and incorrect tags, the terms linear in $r'$
that appear in the observables of the asymmetry are
\begin{equation}
\label{eqn:rpe1}
  (1-2\omega) r' e^{i\delta'}  =  (1 - \omega) r'_c e^{i \delta'_c}
                              - \omega r'_i e^{i \delta'_i}\,.
\end{equation}
This equation gives effective $r'$ and $\delta'$ parameters
in terms of the mistag fraction $\omega$, effective parameters for
correct tags ($r'_c$ and $\delta'_c$) and incorrect tags ($r'_i$ and
$\delta'_i$).
This implies that, in order to have a completely general parameterization
in the data analysis, each tagging category (kaon, lepton, slow
pion, etc.) must have different effective $r'$ and $\delta'$ parameters,
and thus different $b$ and $c$ parameters,
due to the dependence on the mistag fraction $\omega$.
One particular case that is relevant for a kaon tag category is when
$r'_i=0$.
In this case $r' = r'_c (1-\omega)/(1-2\omega)$, which means that the
effective $r'$ is enhanced by a factor of $(1-\omega)/(1-2\omega)$.

The experimental knowledge of $\delta$ depends on $c$, so
even though the $a$ parameter does not depend on $r'$ and $\delta'$,
one does not avoid uncertainties due to $r'$ and $\delta'$ in
the analysis.
The best way to reduce this uncertainty is to take advantage of
the fact that lepton tags are immune to the problem ($r'=0$).
If the fit is performed with an independent $c$ coefficient for
lepton tags, $c_{\rm lep}$ combined with the $a$ parameter measured by
all flavor tagging categories will help resolve $\delta$ and thus
$(2\beta+\gamma)$.

\begin{figure*}
\begin{center}
\epsfig{figure=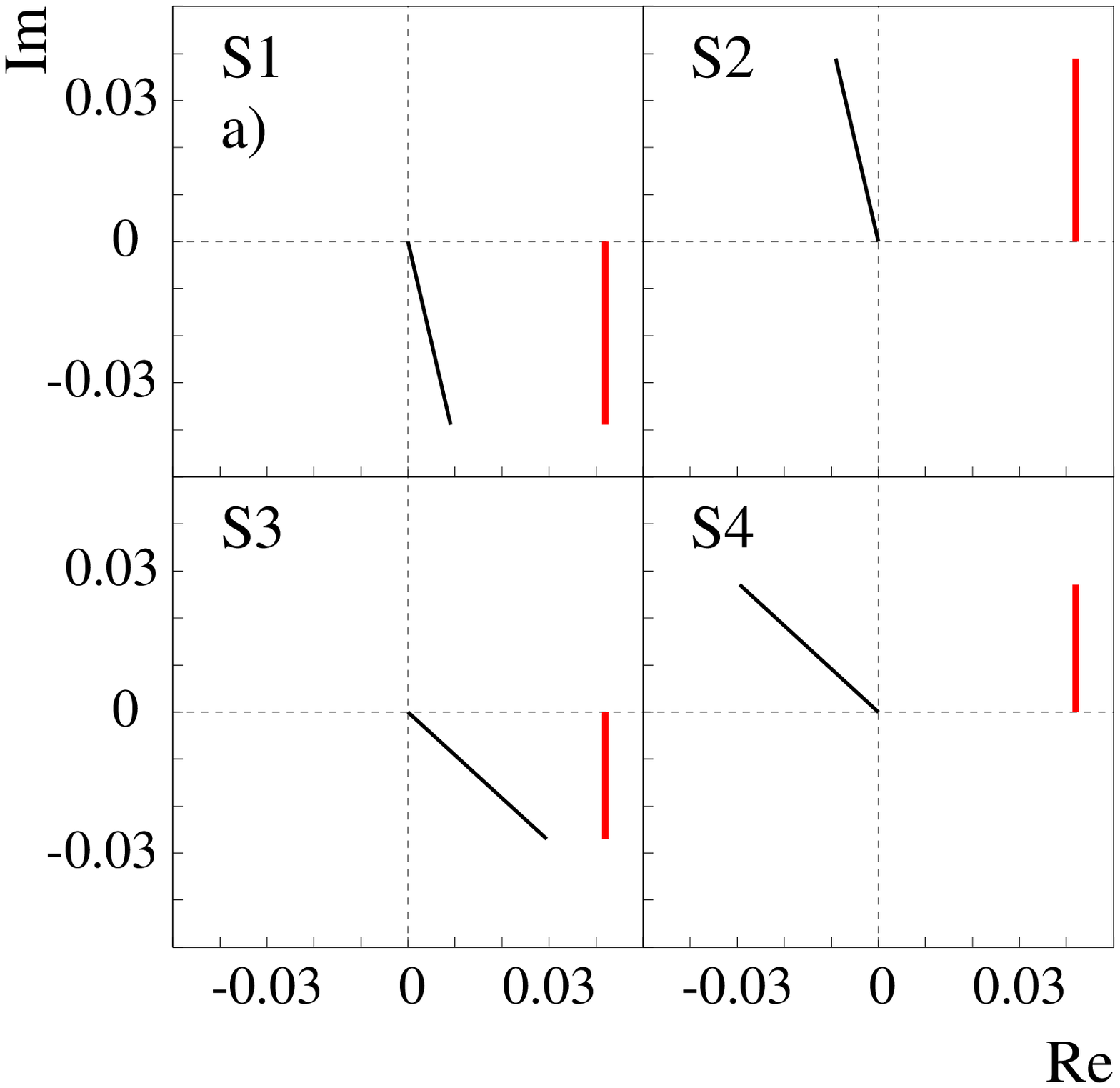,bbllx=0,bblly=15,bburx=490,bbury=490,width=0.325\linewidth}
\epsfig{figure=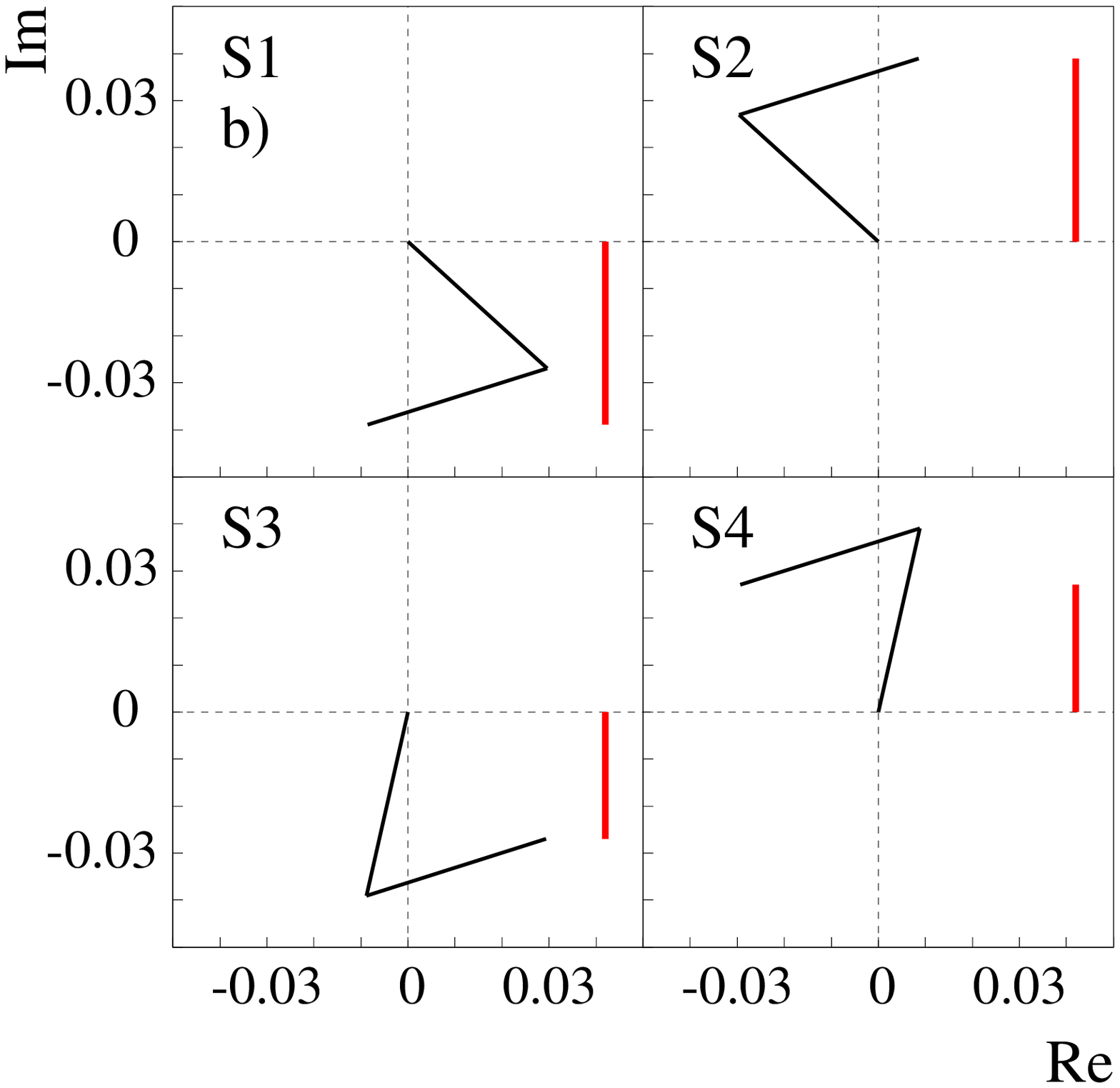,bbllx=0,bblly=15,bburx=490,bbury=490,width=0.325\linewidth}
\epsfig{figure=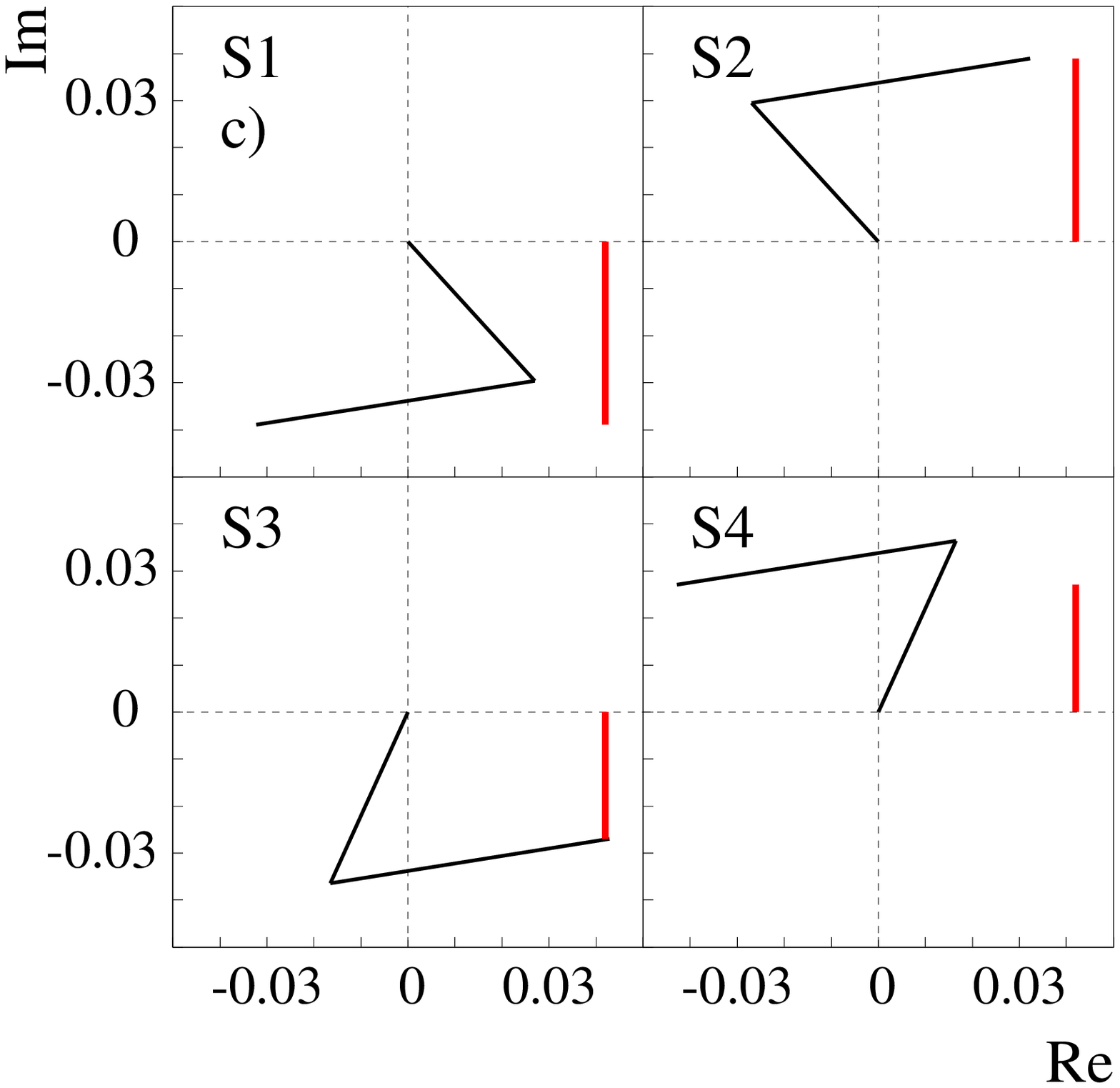,bbllx=0,bblly=15,bburx=490,bbury=490,width=0.325\linewidth}
\caption{
An example of three amplitude configurations that all give the same
set of $S$ coefficients.
For each set (a, b, or c) the $S$ coefficients are consistent
with no tag-side interference ($r'=0$), while this is only true in the first case.
Each configuration is represented by four diagrams showing
the addition of the reconstructed and tag-side amplitude vectors in the
complex plane.
The observable coefficient ($S$) is the imaginary part, represented by the
vertical band on the right side of each diagram.
The parameters for the three configurations are: $r=0.02$; $(2\beta+\gamma)=2.10,\: 1.87,\: 1.73$;
$r'=0.00,\: 0.02,\: 0.03$; $\delta=0.30,\: -0.53,\: -0.58$; and $\delta'={\rm NA},\: 1.57,\: 1.57$.
}
\label{fig:bcpsample}
\end{center}
\end{figure*}

If $r'$ and $\delta'$ are not constrained from other measurements, one must
allow for values of $r'$ and $\delta'$ that are consistent with the measured
values of $b$ and $c$.
Since it is possible to have a measured set of $a$, $b$, and $c$
parameters that are consistent with $r'=0$ when $r'\neq0$,
one must always consider all $r'$
values between 0 and $r'_{\rm max}$ consistent with $b$ and $c$,
where $r'_{\rm max}$ is the largest allowed single-final-state value.
This point is illustrated in Figure~\ref{fig:bcpsample}.
The uncertainty on $(2\beta+\gamma)$ due to $r'$ and $\delta'$ is
maximal when $a$ is small.
In this case, the sensitivity to $(2\beta+\gamma)$ is mostly from
the $c$ coefficient and one must rely on flavor tag categories
that are known to have $r'=0$, such as lepton tags.

Using Monte Carlo pseudo-experiments, we perform a simplified study of the impact 
of DCS tag-side interference on a system with only two tagging categories: 
one for unaffected lepton tags, and the other containing kaon tags. 
The significance ratio of both categories is set to $Q_{\rm lep}/Q_{\rm non-lep}=0.6$. 
All tests use the realistic value of $0.02$ for $r$ and $r'$.
Each category shares the same $a$ parameter. The lepton category constrains $c_{\rm lep}$,
and the kaon category fits $b$ and $c$. 
All fit parameters are unbiased, and conform to Gaussian distributions.
Compared to the situation with no DCS contribution, having one tagging category
and identical errors for its two parameters $a$ and $c$, the statistical error 
on $a$ is unchanged, and that on $c_{\rm lep}$ has increased by a ratio 
compatible with $((Q_{\rm non-lep}+Q_{\rm lep})/Q_{\rm lep})^{1/2}=1.6$.
The parameters $a$ and $b$ show a $20\%$ correlation, while all other 
correlations are smaller than $1\%$.

One experimental strategy for reducing the uncertainties due to $r'$ and $\delta'$ 
would be to constrain them by performing a time-dependent analysis
of a flavor-specific final state that has no DCS contribution ($r=0$), such
as $D^{*+}l^-\overline{\nu}$.
For such a final state, the undiluted $b$ coefficient
is the same as for $D^{*+}\pi^-$ and $c$ now has $r=0$.
This information can be used to recover the $(2\beta+\gamma)$ sensitivity
in the $c$ coefficients in the signal sample that was lost due to the
lack of knowledge of $r'$ and $\delta'$.
Another option would be to include in the analysis events for which it
was not possible to determine the flavor of the tag, so-called untagged events.
From Equations~\ref{eq:abcpars-s1} through~\ref{eq:abcpars-s4},
one can see that the untagged $S$ coefficient
for a reconstructed $D^{*-}\pi^+$ ($D^{*+}\pi^-$) is equal
to $S1+S2=b$ ($S3+S4=-b$),
thus untagged events provide a further constraint on $b$.

The measured $a$, $b$, and $c$ coefficients for the various tagging
categories and samples can be combined by forming a $\chi^2$ using
the measured parameters and the inverted covariance matrix.
This assumes that the measurement uncertainties on the $a$, $b$,
and $c$ parameters are Gaussian.
A constraint on $(2\beta+\gamma)$ can be derived from the $\chi^2$
by scanning the $\chi^2$ vs $(2\beta+\gamma)$ where for each
$(2\beta+\gamma)$ value the $\chi^2$ is minimized with respect
to the unknown parameters $\delta$, $\delta'$, and $r'$.
If there are no external constraints on $r'$ and $\delta'$, such
as from the analysis of $D^{*+}l^-\overline{\nu}$ suggested above,
the $b$ and $c$ parameters from non-lepton tags do not provide
much information, since $r'$ must be varied from its minimum value
compatible with $b$ to its maximum
possible value (for example, see Figure~\ref{fig:bcpsample}).
The non-lepton-tag $b$ and $c$ parameters still must be included in the time-dependent 
fit, but they are not very useful in the $\chi^2$ analysis.

\begin{figure*}
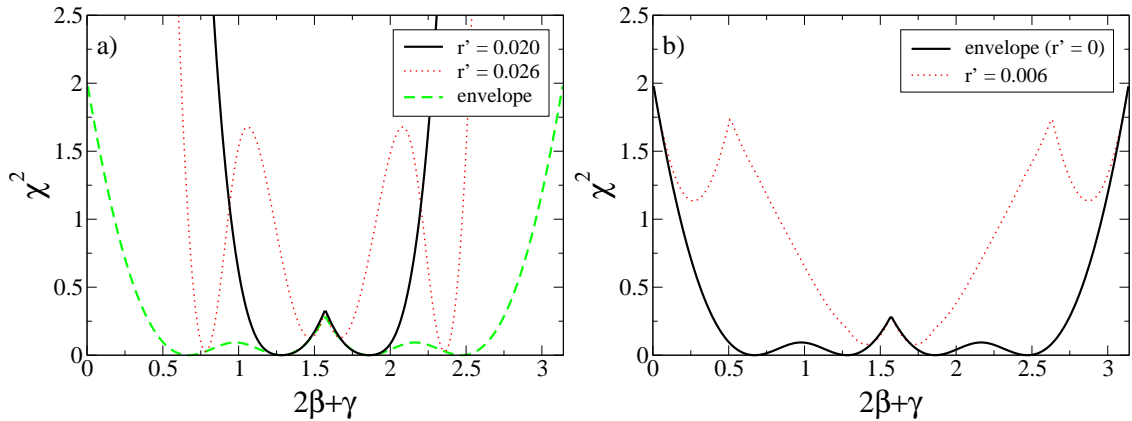

 \begin{center}
 \epsfig{figure=fig5a.eps,width=0.45\linewidth}
 \epsfig{figure=fig5b.eps,width=0.45\linewidth}
 \caption{
    Scans of $\chi^2$ from measured $a$, $b$, and $c$ coefficients
    as a function of $(2\beta+\gamma)$ illustrating two cases:
    a) one with non-zero tag-side DCS interference, and b) one
    without tag-side interference.
    The solid curve in both a) and b) was made with the true value of $r'$.
    The dashed curve labeled 'envelope' in figure a) encloses from below all $\chi^2$ curves
    made with $r'$ ranging from zero to arbitrarily large values.
    The envelope curve in b) coincides with the $\chi^2$ curve made with the true
    $r'$ value of zero.
    The dotted curves in both a) and b) illustrate $\chi^2$ curves with incorrect
    values of $r'$.
    The input values were $(2\beta+\gamma)=1.86$, $r=0.02$, and $\delta=0.9$.
    The tag-side parameters for a) were $r'=0.02$ and $\delta'=0.3$.
    The measured values of the $a$, $b$, and $c$ coefficients were set to the correct values.
    The statistics of the hypothetical measurement correspond
    to roughly 450~fb$^{-1}$ of $B$ factory data from one experiment,
    including a constraint from $D^*l\nu$.
    There is a discreet ambiguity that gives exactly
    the same curves after adding $\pi$ to the horizontal axis
    $(2\beta+\gamma)$.
 }
 \label{fig:chisqscans}
 \end{center}
 \end{figure*}

Figure~\ref{fig:chisqscans} shows an example of the $\chi^2$ procedure
for a hypothetical measurement where $(2\beta+\gamma)=1.86$,
$r=0.02$, and $\delta=0.9$.
The measured values of the $a$, $b$, and $c$ coefficients
were set to the correct values, so the $\chi^2$
is zero at the correct and degenerate solutions.
The two plots in Figure~\ref{fig:chisqscans} illustrate two cases:
a) one with non-zero tag-side DCS interference, and b) one without tag-side
interference.
In addition to the curve which allows for any value of $r'$, labeled
`envelope', additional curves with fixed values of $r'$ are included.
The statistical errors correspond to a measurement from $D^*\pi$
in roughly 450 fb$^{-1}$ of $B$-factory data from one experiment
including a constraint from $D^*l\nu$.

Three important conclusions can be drawn from Figure~\ref{fig:chisqscans}.
First, comparing envelope curves for the $r'=0.02$ case a) to the $r'=0$ case b),
the measurements give nearly identical constraints on $(2\beta+\gamma)$.
This means that the uncertainty on $r'$ and $\delta'$ does not
affect the measurement.
The only degradation with respect to the situation with no tag-side
interference is that, when not including $D^{*+}l^{-}\overline{\nu}$
in the analysis, the non-lepton-tag $c$ parapeters no longer contain
useful information.

The second conclusion is that if $r'$ is non-zero, the constraint on
$(2\beta+\gamma)$ can be better than the case where $r'$ is zero.
If so, the $b$ and $c$ parameters in the $D^{*}l\nu$ sample will in
general be non-zero, and one effectively adds a measurement of
$(2\beta+\gamma)$ from the tag-side $B$.
This can be seen most clearly from the symmetry between the tag-side and
reconstruction-side within the definitions of $a$, $b$, and $c$ in
Eqns.~(\ref{eqn:a}-\ref{eqn:c}).
An extreme, unrealistic example is given by the solid line in case
a) of Figure 5. It shows what the constraint looks like if $r'$ would
equal $0.02$, and if that information were known precisely and
were included in the analysis.

Thirdly, the result for $D^{*}\pi$ alone, after varying $r'$ to arbitrarily
large values, is equivalent to the $\chi^2$ curve constructed from only $a$ and $c_{\rm lep}$.
In other words, when not including the $D^{*}l\nu$ sample in the analysis,
the $b$ and non-lepton-tag $c$ parameters do not contribute to the sensitivity to
$(2\beta+\gamma)$. Again, however, these degrees of freedom must still be included in
the data analysis.
%

\section{Conclusions}
\label{sec:conclusions}

Interference effects between CKM-favored $b\rightarrow c\overline{u}d$
and doubly-CKM-suppressed $\overline{b}\rightarrow \overline{u} c \overline{d}$
amplitudes in final states used for flavor tagging in coherent
$\BzBzb$ pairs from $\Upsilon$(4S) decays introduce deviations from
the standard time evolution assumed in $CP$ violation measurements
at the asymmetric-energy $B$ factories.
To our knowledge, the uncertainty introduced by this interference has been neglected in most
$B$ factory $CP$ violation measurements published to date, with the exception
of~\cite{babarS2B}.
The uncertainties introduced in the $\sin2\beta$ measurement in $(c\overline{c})K^0$
decay modes and the time dependent analysis of the $\pi^+\pi^-$ final state
are at most of the order of 5\% and can be limited to $<2$\% in most
cases with reasonable assumptions.

In proposed measurements of $\sin(2\beta+\gamma)$ which explicitly use
interference between CKM-favored and doubly-CKM-suppressed
amplitude contributions in the final state that is reconstructed,
such as $D^*\pi$, tag-side interference effects can be as large as
the interference effects one is trying to measure.
In any such analysis, the data must be analyzed in a way that is
general enough to allow for tag-side interference effects.
We have proposed a general framework for dealing with tag-side interference effects
in $\sin(2\beta+\gamma)$ measurements.
It is possible to achieve an experimental sensitivity to $(2\beta+\gamma)$ similar
to the originally proposed measurements.

\begin{acknowledgments}
We would like to thank Pat Burchat for helpful discussions.
The work of O.L., R.C, and D.K. was supported by the U.S. Department of Energy under
contracts DE-FG03-91ER40618, DE-AC03-76SF00098, and DE-FG03-91ER40679, respectively.
The work of M.B. was supported by F.O.M. program 23 (The Netherlands).
\end{acknowledgments}


\end{document}